\documentstyle{amsppt}
\NoBlackBoxes
\NoRunningHeads
\topmatter
\title Boundary currents and hamiltonian quantization of fermions in
 background fields \endtitle
 \author Jouko Mickelsson \endauthor
 \affil Theoretical Physics, Royal Institute of Technology, SE-10044,
 Stockholm, Sweden. e-mail jouko\@theophys.kth.se \endaffil
 \endtopmatter

\document
\baselineskip=18pt
ABSTRACT The current algebra generated by fermions coupled to external
gauge potentials and metrics on a manifold with boundary is discussed.
It is shown that the previous methods, based on index theory arguments and
used in the case without boundaries, carry over to the present problem.
The resulting current algebra is the same as obtained from a quantization
of bosonic Chern-Simons theories on  space-times with nonempty boundaries.
This means that also in the fermionic setting the construction is 'holographic',
the Schwinger terms reside on the boundary.

\vskip 0.6in
    \define\tr{\text{tr}}
                                       \define\Di{\text{\it Diff\rm}}
     \redefine\l{\lambda}
The current algebra on manifolds with boundary has drawn recently great
attention because of the correspondence between gravitation theories
on anti- de Sitter type (odd dimensional) space-times and conformal field
theory on the even dimensional boundary, [1,2]. In hamiltonian formulation one
tries to construct the theory from the current algebra (for fixed time)
on the odd dimensional boundary of the even dimensonal space. In presence
of gauge symmetries a prototype for the current algebra is an affine
Lie algebra on the boundary in the case of a $2+1$ dimensional
interior space-time. This arises for example if the action functional
(in the bulk) contains Yang-Mills and Chern-Simons terms, [4]. The
boundary currents play also an important role in discussions of
the quantum Hall effect, [3].

The purpose of the present letter is to show how the boundary current algebra
arises from the canonical hamiltonian quantization of fermions in background
gauge fields.
The method is closely related to previous studies of quantization
of fermions in background gauge fields when the physical space-time is
even dimensional and without boundary, [5,6], using a mixture of index theory
and a variant of theory of gerbes.

Consider first an even dimensional compact Riemannian manifold $M$ without
boundary. We
assume that $M$ is oriented and with a fixed spin structure. Let $V$
be a trivial vector bundle over $M$ with an unitary action of a gauge
group $G.$ Let $S$ be the tensor product of $V$ and the Dirac spinor bundle
over $M.$
Given a gauge potential $A$ in $V$ and a metric $g$ on $M$
we have a Dirac operator $D_{g,A}$ acting on the smooth sections of $S.$
The Dirac operator on a compact manifold is
Fredholm. We denote by $H=H(g,A)$ the Hilbert space of square integrable
sections of the bundle $S.$ We denote by $\Cal A$ the space of all smooth
vector potentials
(with values in the Lie algebra $\bold g$ of the gauge group) and by
$\Cal M$ the space of Riemann metrics on $M.$

We recall the construction of the Dirac determinant bundle (for a more
detailed discussion and references to the extensive original literature, see
[7, Chapter 9]). Let $\lambda\in\Bbb R$ and $U_{\l}$
the set of potentials and metrics such that $\l$ is not in the spectrum of
$D_{g,A}.$ Then
$U_{\l}\subset \Cal M\times \Cal A=B$ is an open set and over $U_{\l}$ we
define a complex line bundle $DET_{\l}$ such that the line at $z\in U_{\l}$ is
$$DET_{\l}(z) = \wedge^{top}(S^+_{\l}) \otimes \wedge^{top}(S^-_{\l})^*,
\tag1$$
where $S^{\pm}_{\l}$ is the space of positive (negative) chirality modes
in the spectral subspace $(D_z)^2 < \l.$

Let $0< \lambda < \lambda'.$ We define $DET_{\l\l'}$ to be a trivial complex
line bundle over the intersection $U_{\l\l'}= U_{\l} \cap U_{\l'}.$ The
line at $z$ is the tensor product
$$DET_{\l\l'}(z) = \wedge^{top}(S^+_{\l\l'}) \otimes \wedge^{top}
(S^-_{\l\l'})^*,\tag2$$
where $S^{\pm}_{\l\l'} = (S^{\pm}_{\l})^{\perp} \cap S^{\pm}_{\l'}.$
Clearly $DET_{\l'}= DET_{\l\l'} \otimes DET_{\l}$ over $U_{\l\l'}.$
Furthermore, $DET_{\l\l'}$ has a canonical trivialization defined as follows.
Let $v_1,v_2,\dots, v_n$ be any orthonormal basis of $S^+_{\l\l'}.$ Then
the vectors $J\epsilon v_i$ form a basis of $(S^-_{\l\l'})^*,$ where $\epsilon
=D_z/|D_z|$ and $J$ is the canonical antilinear isomorphism from a Hilbert
space to its dual. Note here that $D_z$ anticommutes with the chirality
operator whereas $|D_z|$ commutes with chirality and thus $\epsilon$ maps
$S^+$ to $S^-.$   As a consequence of eq. 2 and the triviality of the bundles
$DET_{\l\l'},$ the local line bundles $DET_{\l}$ patch up to a globally
defined line bundle $DET$ over $B.$ The construction is gauge equivariant,
leading to a line bundle $DET$ over $B/(\Di_0(M)\times\Cal G_0).$
Here $\Cal G_0$
is the group of based gauge transformations, i.e., those gauge transformations
which at a fixed base point $p\in M$ take the value $1.$ $\Di_0(M)$ is the
group of diffeomorphims $h$ such that $h(p)=p$ and the derivative of $h$
at $p$ is the identity.

If the Dirac operator $D_z$ does not have zero modes then the
construction of the fermionic Fock space and the vacuum is canonical.
The representation of the CAR algebra is fixed by the polarization
$H(z)= H_+(z) \oplus H_-(z)$ of the one-particle Hilbert space to positive
and negative spectral subspaces. The vacuum is characterized by the
property
$$a^*(u)|z>=0= a(v)|z>$$
for all $u\in H_-$ and $v\in H_+.$ The only nonzero anticommutators are
$[a^*(u), a(v)]_+ = \ <v,u>,$ the inner product
in $H$ being antilinear in the first argument.

In the case there are zero modes
there is an ambiguity. One has to decide which of the zero modes should
belong to $H_+$ and which to $H_-.$ The choice cannot be made in an arbitrary
manner if we want to have a smooth bundle of Fock spaces.
We resolve this ambiguity by requiring
that the vacuum is annihilated by all $a^*(v)$'s for positive chirality
zero modes and all $a(u)$'s corresponding to negative chirality zero
modes. That is, we may view the vacuum as an element in the line
$$DET=\wedge^{top}(S^+_{D_z=0}) \times \wedge^{top}(S^-_{D_z=0})^*$$
belonging to
the  space $\wedge(S^+_{D_z=0}) \otimes \wedge(S^-_{D_z=0})^*.$

To be more precise
and in order to show the smoothness of this construction one can proceed
as before in the case of the bundle $DET.$ First define local Fock bundles
$\Cal F_{\l}$ over $U_{\l},$ with $\l >0,$ as follows.
Let $H_{+,\lambda}$ (resp.
$H_{-,\lambda}$) be the spectral subspace $D_z <\lambda$ (resp. $D_z <
-\lambda$). Set
$$\Cal F_{\l}(z) = \wedge (H_{+,\l} \oplus S^+_{\l}) \otimes \wedge
(H_{-, \l} \oplus S^-_{\l})^*.\tag3$$
The vacuum $|z,\l>$ in $\Cal F_{\l}(z)$ is defined as a normalized element
in $DET_{\l}(z) \subset \Cal F_{\l}(z).$ Wedging this with the canonical
element in $DET_{\l\l'}$ we obtain an element in $DET_{\l'}(z)\subset
\Cal F_{\l'}(z).$ This provides us an identification of $\Cal F_{\l}$
with $\Cal F_{\l'}.$ Thus we obtain a globally defined bundle of Fock spaces
$\Cal F$ over $B.$

The creation and annihilation operators are defined in a Fock space $\Cal F=
\wedge(W) \otimes \wedge(W^{\perp})^*$ in the standard way. The creation operator
$a^*(u)$ for $u\in W$ is represented by the
exterior product $u\wedge$ and the annihilation by the contraction
$i_u$ (which is antilinear in $u$). The creation and annihilation operators for
$v\in W^{\perp}$ are  $a^*(v) = i_{Jv}$
and the annihilation operator $a(v)= Jv \wedge.$

In the case of a manifold with boundary there are important modifications to
the construction above. We have to specify elliptic boundary conditions for the
Dirac operators $D_z.$

We first consider the APS boundary
conditions, [8], defined by $\psi|_{\partial M} \in P_{\l}(H^b)$ over those
parameters $z=(g,A)$ such that $\lambda\notin Spec(h_z).$ Here $P_{\l}$
is the spectral projection to the part of the boundary Hilbert space $H^b$
where $h_z < \l;$ $h_z$ is the boundary Dirac operator to be defined below.
In order to apply
the APS theory we have to assume that the geometry is tubular near the
boundary. This means that the metric is a product metric near $\partial M$
and the gauge potentials have vanishing normal derivatives at the boundary.
The Dirac equation near $\partial M$ can be written as
$$\frac{\partial}{\partial t} \psi = h_z \psi$$
where $t$ is a coordinate in the normal direction near the boundary,
$h_z$ is the (hermitean) boundary operator $\gamma_t \gamma^i \nabla_i,$
$i=1,2,\dots ,2n-1$ and $\nabla_i$ denotes the combined gauge and  Levi-Civita
covariant derivative in the $i$:th local basis vector direction.

The operator $h_z$ commutes with the chirality operator $\gamma_{2n+1}$ and
therefore the positive and
negative eigenvalues for the Dirac operator in the bulk come in pairs like
in the case without boundary and the determinant line bundle and the Fock
spaces are defined as before. The crucial difference is in the fact that
the determinant line bundle (and therefore also the Fock spaces) are only
defined locally, over the sets $V_{\lambda}=\{z\in B| \l\notin Spec(h_z)\},$
in the parameter space.

In order to define the Fock spaces globally one can choose a spectral
section on the boundary (for an alternative construction see [6]).
We denote by $DET(D,P)$ the determinant bundle for the family $D_z$ of
Dirac operators with the boundary conditions $\psi|_{\partial M} \in
P(H^b(z))$. Here $P=P^{(z)}$ is a projection operator acting in the boundary
Hilbert space $H^b(z)$. It should depend smoothly on the parameter $z$ and
the difference $P - P_{\l}$ must be Hilbert-Schmidt.
The determinant bundle in the bulk becomes then
$$DET(D,P) = DET(D, P_{\lambda}) \otimes DET_F(P^+_{\lambda}, P^+)\tag4$$
where $P$ is the spectral section, [10].
We have set $P^+=\pi_+ P$ where $\pi_{\pm}$ are the chiral
projectors on the boundary.
Since $P^{(z)}$ and $P_{\l}(z)$ are Hilbert-Schmidt related, the
latter line bundle $DET(P^+_{\l},P^+)$ can be taken as the (relative) Fredholm
determinant bundle over a restricted Grassmannian manifold as discussed in
[11]. We have also performed the standard identification of $DET(D)$ as
the tensor product $\wedge^{top}(ker D^+) \otimes \wedge^{top}(coker D^+)^*$
which leads to the localization of the domain to the positive chirality
sector in (4).

The gauge and diffeomorphism group action on the first factor in the above
formula (4) is equivariant but not on the second factor because in general there
is no smooth equivariant choice for the spectral section $P.$
It follows that there is a potential obstruction to lifting the group
actions to the total space of the Fock bundle due to a nontrivial extension
coming from the lifting of the group actions on potentials and metrics to
the complex line bundle $DET_F(P^+_{\lambda}, P^+).$
On the level of Lie algebras
this obstruction is seen as nontrivial Schwinger terms in current commutation
relations.  As explained in [6] the Schwinger terms can be computed from
the curvature of the determinant bundles. APS index theory determines the
cohomology class of the curvature form.

Any parameter $z\in B$ determines a boundary geometry $z_b.$ On the other hand,
for any $z_b$ we can choose a continuation to a metric
and gauge connection $\hat z$ in the bulk. The map $z\mapsto \hat z$ clearly
defines a map from $B/(\Di_0 \times \Cal G_0)$ to a contractible set of
parameters $\hat z$ (since the space of the parameters $z_b$ is contractible).
Let $1-P^{(z)}$ be the \it Calderon projection \rm associated to the Dirac
operator $D_{\hat z}.$ By Calderon projection we mean the projection to the
space of boundary values of solutions of the Dirac equation $D_{\hat z}\psi
=0.$ It is known that $P^{(z)}$ differs by a Hilbert-Schmidt operator (in fact,
by a smoothing operator) from the spectral projection $P_{\lambda}$ for
the boundary hamiltonian; if we drop the requirement concerning the vanishing
of the normal derivatives of metrics and potentials at the boundary,
one has more general $L_p$ type estimates replacing the Hilbert-Schmidt
condition, [12].
It follows that we may choose $P_z$ as the
spectral section above. Note that now everything in the parametrization
of the determinant lines in $DET(P^+_{\l}, P^+)$ \it depends on the boundary
data only. \rm

The curvature of $DET(P^+_{\l},P^+)$ can be computed from the APS theory.
For determining the Schwinger terms one needs to know the curvature
to the gauge directions only, [6]. The result is equivalent to a
standard local quantity; when the boundary is 1-dimensional one gets
the central term in an affine Lie algebra and in the 3-dimensional case
it is extension part in the Mickelsson-Faddeev algebra.

Following [9] we can construct explicitly the vacuum bundle, that is, the
chiral
Dirac determinant bundle from the chiral anomaly functional. Restricting to the
case of a fixed metric and $\partial M=\emptyset$
the sections of the bundle $DET$ are complex functions
$\psi:\Cal A \to \Bbb C$ such that
$$\psi(A^g) = \psi(A) \cdot e^{i\omega(g;A)}\tag5$$
where $A^g=g^{-1}Ag+g^{-1}dg$ and the anomaly functional $\omega$ is an
integral of a local
differential polynomial in $A$ and $dg g^{-1}.$ It satisfies the cocycle
property
$$\omega(g_1g_2;A)= \omega(g_1;A) +\omega(g_2; A^{g_1}).\tag6$$

The important property of this construction is that it extends to the case
with boundary. The condition (5) should now hold for all potentials $A$
which obey the tubularity condition at the boundary and for those gauge transformations
$g$ which smoothly approach the value $1$ at the boundary. Let $\Cal G_{00}$
be the group of gauge transformations with vanishing normal derivatives at the
boundary and which are equal to $1$ on the boundary. We can
mod out by $\Cal G_{00}$ and we obtain a determinant
bundle over $\Cal A/\Cal G_{00}.$
We have not specified explicitly the boundary conditions $P$ as a function of
the boundary data. The reason
is that \it whatever choice we make for the boundary projections \rm
$P$ (satisfying
the Hilbert-Schmidt relations and depending smoothly on boundary data)
\it the resulting bundles will be equivalent. \rm
This is seen from (4). If we change the boundary conditions then the
corresponding determinant bundles are related by the factor
$DET(P^+,{P'}^+).$ Modding out by the gauge and diffeomorphism symmetries in the
bulk does not affect the boundary data. So even after pushing things to
$\Cal A/\Cal G_{00}$ the bundle $DET(P^+,{P'}^+)$ will be a bundle over the
affine space
of all boundary potentials and therefore automatically trivial. Clearly,
the same comments apply to metrics and diffeomorphisms.

The bigger group $Map(M,G)$ of all gauge transformations (including also those
which are not equal to $1$ at the boundary but still have vanishing normal
derivatives)  acts projectivly in the space of
sections $\Gamma(DET)$ through the formula
$$(T(g)\psi)(A) = e^{i\gamma(g;A)} \psi(A^g),$$
where $\gamma$ is an integral of another local differential polynomial,
[9]. This functional does not satisfy the cocycle condition. This means
that $T(g_1)T(g_2)= T(g_1g_2) e^{i\omega_2(g_1,g_2;A)}$ for some 2-cocycle
$\omega_2.$ Infinitesimally this means that the Lie algebra of $Map(M,G)$ is
extended by an abelian ideal consisting of functions $f:\Cal A\to \Bbb C,$
$$[(X,f),(Y,g)] =([X,Y], X\cdot g-Y\cdot f +c(X,Y;A)),\tag7$$
where $c$ is a Lie algebra cocycle and the action $X\cdot g$ is the Lie
derivative of a function determined by the infinitesimal gauge transformation
$\delta_X A= [A,X] +dX.$

As promised in the abstract, a basic property of the functional $c(X,Y;A)$ is
that it depends only on the boundary values of the fields, [9].
In $2+1$ dimensions we have
$$c(X,Y;A) = \frac{i}{2\pi}\int_{M_2} \tr \, dX\wedge dY = \frac{i}{2\pi}
\int_{\partial M_2} \tr \, XdY,\tag8$$
where $M_2$ is the 2-dimensonal physical space. In $4+1$ dimensions we have
$$\align c(X,Y;A)&= \frac{i}{24\pi^2} \int_{M_4} \tr\, dA\wedge(dX\wedge dY-
dY\wedge dX)\\
&= \frac{i}{24\pi^2} \int_{\partial M_4} \tr \, A\wedge (dX\wedge dY-dY
\wedge dX).\tag9\endalign$$

\newpage
\bf References \rm \newline\newline

[1] J. Maldacena: The large $N$ limit of superconformal field theories
and supergravities. Adv. Theor. Math. Phys. \bf 2, \rm 231 (1998).
hep-th/9711200

[2] E. Witten: Anti-de Sitter space and holography. Adv. Theor. Math.
Phys. \bf 2, \rm 253 (1998). hep-th/9802150

[3] J. Fr\"ohlich and U.M. Studer: Gauge invariance and current algebra
in nonrelativistic many-body theory. Rev. Modern Phys. \bf 65, \rm
no. 3, part 1, 753 (1993)

[4] J. Mickelsson: On a relation between massive Yang-Mills theories
 and dual string models. Lett. Math. Phys. \bf 7, \rm 45 (1983)

[5] J. Mickelsson: On the hamiltonian approach to commutator anomalies
in $3+1$ dimensions. Phys. Lett. \bf B241, \rm 70 (1990)

[6] A. Carey, J. Mickelsson, and M. Murray: Index theory, gerbes, and
hamiltonian
quantization. Commun. Math. Phys. \bf 183, \rm 707 (1997); Bundle gerbes
applied to quantum field theory. hep-th/9711133, to be publ. in Rev. Math.
Phys.

[7] N. Berline, E. Getzler, and M. Vergne: \it Heat Kernels and Dirac Operators.
\rm Grundlehren der mathematischen Wissenschaften 298. Springer-Verlag  (1992)

[8] M.F.  Atiyah, V.K. Patodi, and I.M. Singer: Spectral asymmetry
and Riemannian
geometry I-III. Math. Proc. Camb. Phil. Soc. \bf 77, \rm 43 (1975); \bf 78,
\rm 405 (1975); \bf 79, \rm 71 (1976).

[9] J. Mickelsson: Kac-Moody groups, topology of the Dirac determinant bundle
and fermionization.
Commun. Math. Phys. \bf 110, \rm 173 (1987).
\it Current Algebras and Groups. \rm Plenum Press,
London and New York (1989)

[10]  S. Scott: Splitting the curvature of the determinant line bundle.
math.AP/9812124. Proc. Amer. Math. Soc. (to appear)

[11] A. Pressley and G. Segal: \it Loop Groups. \rm Clarendon Press,
Oxford (1986)

[12] L. Fiedlander and A. Schwarz: Grassmannian and elliptic operators.
funct-anal/9704003

\enddocument